\documentstyle[12pt]{article}
\begin {document}
\begin{flushright} {	OITS-597}\\
February 1996
\end{flushright}
\vskip.7cm

\begin{center} {\Large {\bf Quark-hadron phase transition with \\
surface fluctuation}}
\vskip .75cm
 {\large{\bf  Zhen Cao and Rudolph C. Hwa} }
\vskip.5cm
 {Institute of Theoretical Science and Department of Physics\\  University of
Oregon, Eugene, OR 97403}
\end{center}

\vskip .5cm

\begin{abstract}
The effect of surface fluctuation on the observables of quark-hadron phase transition
is studied.  The Ginzburg-Landau formalism is extended by the inclusion of an extra
term in the free energy that depends on the vertical displacements from a flat
surface.  The probability that a bin has a particular net displacement is determined
by lattice simulation, where the physics input is color confinement.  The surface
fluctuation from bin to bin is related to multiplicity fluctuation, which in turn is
measured by the factorial moments.  It is found that both the $F$-scaling behavior and
the scaling exponent are essentially unaffected by the inclusion of surface
fluctuation.
\end{abstract}

\section{Introduction}

Scaling behaviors of hadronic observables in quark-hadron transition (PT) have been
studied in the framework of Ginzburg-Landau (GL) theory
\cite{1}-\cite{4}.  Since it is a mean-field theory, it cannot account for
the dynamical fluctuations that can occur in heavy-ion collisions.  An
attempt to amend that defect has been made by treating the problem in the
Ising model, so that by lattice simulation the effect of spatial
fluctuation of hadron density at the surface of quark-gluon plasma can be
considered \cite{5}.  In this paper we treat the problem of spatial
fluctuation in a very different way.  We generalize the GL description of
second-order PT in
$2D$ by including a term that accounts for surface fluctuation.  The
deviation from a flat surface is simulated by use of a term in the free
energy that represents the confining potential between quarks.  By
associating bulges on the surface to regions where hadronization can take
place more readily, we can determine the effect of spatial fluctuations
on the hadronic observables.  Thus this work represents a significant
step toward making the GL formalism a more realistic description of the
PT problem in heavy-ion collisions.

In Refs. \cite{1}-\cite{5} it has been found that the scaling behavior
characteristic of the PT is associated with the normalized factorial
moments $F_q$.  Actually, that scaling behavior does not exhibit
intermittency \cite{6}, which is the power law
$F_q \propto M^{\varphi_q}$, where $M$ is the number of bins in a fixed volume of
phase space.  Instead of $M$ scaling, which we use to refer to the
$M^{\varphi_q}$ behavior, we have found that pure GL leads to $F$ scaling,
which is 
\begin{eqnarray}
F_q \propto F_2^{\beta_q} ,
\label{1}
\end{eqnarray}   
when $M$ is varied.  Furthermore, $\beta_q$ satisfies the simple formula \cite{1}
\begin{eqnarray}
\beta_q =(q-1)^\nu  .
\label{2}
\end{eqnarray}
It is this $\nu$ that has been referred to as the scaling exponent.  It is not to
be confused with the conventional critical indices in statistical physics, to
which $\nu$ bears absolutely no relationship.  In problems where the temperature
$T$ is not observable, but the multiplicities of particles produced are (such as
photons in lasers \cite{7} and pions in heavy-ion collisions \cite{4}), $F$
scaling has been found to be the only scaling behavior that can be put to
experimental tests.  Thus far the value of $\nu$ at $1.3$ has proven to be
remarkably universal theoretically, and has been verified experimentally to be
accurate in quantum optics \cite{7}.  In this paper we shall check whether $F$
scaling remains valid when spatial fluctuations are taken into account, and if so,
whether $\nu =1.3$ is still true.

\section{Formulation of the problem}

We give first a very brief review of the usual method in the GL approach to
hadronic observables \cite{1}-\cite{4}.  The multiplicity distribution
$P_n$ is expressed as a functional integral
\begin{eqnarray}
P_n = {1 \over Z} \int {\cal D} \phi {1 \over n!} \left(\int d^2 x |\phi|^2\right)^n
e^{-\int d^2 x |\phi|^2} e^{-F}
\label{3}
\end{eqnarray}
where $Z= \int {\cal D} \phi\,  e^{-F}$, and the GL free energy is 
\begin{eqnarray}
F= \int d^2 x \, (a |\phi|^2 + b |\phi|^4 + c |\nabla \phi|^2).
\label{4}
\end{eqnarray}
Here the spatial integration is over a $2D$ space, which may be taken to be the
surface of the cylinder containing quark-gluon plasma.  When the surface
temperature $T$, on which the parameters $a$, $b$, and $c$ in (\ref{4}) depend, is
low enough, hadrons are formed on that surface and are removed from the confining
medium that exists within.  If the integration in $\vec x$ is limited to a small
bin of size $A$, then the $A$-dependence of $P_n$ can best be determined in terms
of $F_q$, where 
\begin{eqnarray}
F_q ={\left<n(n-1) \cdot \cdot \cdot  (n-q+1)\right> \over \left<n\right>^q}
\label{5}
\end{eqnarray}
the average being defined by use of the distribution $P_n$.  It is this $F_q$ that
has the scaling properties (\ref{1}) and (\ref{2}).

In most treatments \cite{1}-\cite{4} the third term in (\ref{4}) is
neglected.  Its inclusion is considered by Hwa and Pan \cite{2}, and is
found to lead to only a small change in the value of $\nu$.  Our approach
here is to replace that term by another that has explicit spatial
dependence with dynamical origin.  But before embarking on that, let us
rewrite the formalism in a more compact way.

If we write (\ref{5}) in the form
\begin{eqnarray}
F_q =f_q/f_1^q ,
\label{6}
\end{eqnarray}  
then we have
\begin{eqnarray}
f_q =\sum_{n=q}^{\infty} {n! \over (n-q)!} P_n= \int dm\  m^q D_q (m), 
\label{7}
\end{eqnarray}
where $D_q (m)$ is proportional to $e^{-F(m)}$, and 
\begin{eqnarray}
m= \int_A d^2 x \left|\phi (x) \right|^2 =A |\phi|^2 .
\label{8}
\end{eqnarray}
Without the last term of (\ref{4}), $\phi (x)$ may be regarded as constant inside
$A$ so the last expression in (\ref{8}) follows.  With the interpretation of
$|\phi|^2$ as the hadron density, an order parameter, we may regard $m$ as the
``dynamical'' multiplicity in a bin of area $A$.  The statistical fluctuation
represented by the Poissonian factors in front of $e^{-F}$ in (\ref{3}) is removed
by taking the factorial moment, so that the series in terms of the
``experimental'' multiplicity $n$ in (\ref{7}) is replaced by the integral over
the ``dynamical'' multiplicity $m$, which need not be an integer.  $f_q$ has
multiplicative factors that are cancelled by the same factors in $f_1^q$, when
the ratio in (\ref{6}) is taken.  Leaving out those factors, we can write $f_q$
more transparently as
\begin{eqnarray}
f_q =\int_{0}^{\infty} dm \ m^q {\rm exp}[-(m-m_a)^2 / (A/b)] ,
\label{9}
\end{eqnarray}
where
\begin{eqnarray}
m_a =-Aa/2b .
\label{10}
\end{eqnarray}
PT occurs when $a<0$, so $m_a$ is always positive in the following consideration. 
Note that $a$ is dimensionless, while $b$ has the dimension of area.  Thus the
physical $A$ always appears in the ratio $A/b$, i.e., in the GL
description the area is measured in units of $b$.  In \cite{1} it is
shown that $F_q$ is a function of only one composite variable,
$Aa^2/2b$.  In the following because of our way of introducing spatial
fluctuations, there will be seperate dependences on
$a$ and $A/b$.  Hence, (\ref{9}) will be our starting point, when there is no
spatial fluctuation.

To introduce surface fluctuations, let us first map the $2D$ surface to a square
lattice of size $L^2$, on which we impose periodic boundary condition.  With bins
of size $\delta^2$, there are $M= (L/ \delta)^2$ bins on the lattice.  At each
lattice site we assign a variable $z_i$ that represents the vertical displacement
from the flat surface at site $i$.  We restrict $z_i$ to be in the range $-1 \leq
z_i \leq 1$, deferring the physical scale of perpendicular displacement to another
parameter to be introduced below.  Zero displacement is defined by the global
constraint
\begin{eqnarray}
\sum_{i \in L^2} z_i=0 ,
\label{11}
\end{eqnarray}
where the sum is over all sites on the lattice.

The physics governing the $z$-displacement is color confinement.   
In the discretization of the physical space we do not mean that the lattice
spacing is to be identified with the actual average distance between partons on the
surface of the plasma.  The lattice is a coarse-grain representation of the
surface, and we let each site carry the net color charge of the basic cell
(of area lattice-spacing squared) surrounding the site. For brevity we call such a
color charge at the site a quark.  Our aim is to simulate vertical fluctuations of the
surface.  With the lattice spacing being fixed, the vertical displacement of a site
relative to its neighbors represents further separation among the quarks.  Since
energy is required to increase the quark separation, we add a term to the free energy
to inhibit vertical displacement, calling it $F_s$ to represent spatial contribution,
\begin{eqnarray}
F_s =C \sum_{\left< ij \right>} \left| z_i - z_j \right| ,
\label{12}
\end{eqnarray}
where the sum is over the nearest neighbors $j$ of each site $i$, and then over
all $i$.  The confinement parameter $C$ includes the $(k_B T)^{-1}$ factor, as in
(\ref{4}), and the scale factor for perpendicular displacement mentioned earlier. 
The $|z_i - z_j|$ dependence represents the linearly increasing portion of the
confinement potential.  The $-1/r$ Coulombic portion is omitted partly because the
lattice spacing imposes a minimum distance between quarks that prevents $r$ from
becoming small, and partly because we want to avoid introducing another
parameter.  In place of the last term of (\ref{4}), $F_s$ as given in (\ref{12})
provides a sensible description of spatial dependence of the energy of the system
that incorporates the color dynamics into the GL formalism.  A key link yet to be
specified is the way in which $z_i$ is related to the order parameter $\phi$ or
the dynamical multiplicity $m$.

Our procedure is to use the Boltzmann factor exp($-F_s$) to simulate a
configuration of $z_i$ throughout the lattice.  Clearly, because of thermal
fluctuation the vertical displacement need not be $z_i =0$ for all $i$, even
though that is the most favored configuration.  Because of (\ref{11}) there
are as many net positive displacements as there are negatives on the lattice. 
However, as we focus on a particular bin of area $\delta^2$, there may be a
nonvanishing displacement for the bin
\begin{eqnarray}
\zeta = \sum_{i \in \delta^2} z_i .
\label{13}
\end{eqnarray}
We identify a positive $\zeta$ as having a net outward protrusion of the surface
at the bin, and a negative $\zeta$ as having a net inward indentation.  Nucleation
dynamics suggests that hadronization is more likely to occur at the location of
protruding surface and less likely at where the surface is indented. 
Thus given a value of $\phi$ for a bin, the dynamical multiplicity should
deviate from $m$ in a way that is proportional to $\zeta$.  That is, we
should consider a convolution $G
\otimes S$, where $G$ is the GL term given by the exponential in (\ref{9}), and $S$
represents the spatial fluctuation specified by $S(\zeta)$, yet to be determined,
which describes the probability that a bin has a net vertical displacement $\zeta$.

More precisely, we upgrade (\ref{9}) to include spatial fluctuations by writing
\begin{eqnarray}
f_q =\int_{0}^{\infty} d\mu \, \mu^q D(\mu) ,
\label{14}
\end{eqnarray}
where the dynamical distribution $D(\mu)$ is 
\begin{eqnarray}
D(\mu) = G \otimes S = \int_{0}^{\infty} dm \, G(m) S(\mu -m)
\label{15}
\end{eqnarray}
and
\begin{eqnarray}
G(m) = {\rm exp}[-(m-m_a)^2 / (A/b)] .
\label{16}
\end{eqnarray}
In the next section we shall determine the distribution in $\zeta$. What remains
to be specified is the relationship between $\zeta$ and hadron multiplicity.
Without getting involved with the details of nucleation, surface tension, etc., let
us introduce a new parameter $\alpha$ that summarizes all the dynamical details
relating vertical surface displacement to multiplicity fluctuation, $\mu -m$, so
that
\begin{eqnarray}
\mu = m+ \alpha \zeta  .
\label{17}
\end{eqnarray}
Recall that $m$ is the mean multiplicity in the GL theory.  The spatial
fluctuation of the multiplicity distribution, $S(\alpha \zeta)$, to be determined
in the next section is what appears in (\ref{15}) for the net dynamical
distribution $D(\mu)$.
\section {Simulation on the lattice}

Our first task is to reproduce to GL result \cite{1} (without surface
fluctuation) by lattice simulation.  In the analytical method used in \cite{1},
there is only one composite variable, $x= Aa^2 /2b$, that $F_q$ depends on.  $F$
scaling is found to hold in the range $0.1< {\rm ln} F_2 <0.44$, which
corresponds roughly to $3< {\rm ln} F_{10} <9$.  From Fig.\ 1 of \cite{1} one
can see that such ranges of $F_q$ are for $x$ in the range $-2< -{\rm ln}\, x
<4$.  Since $x$ is a composite variable, it does not matter how $a$, $b$ and
$A$ are separately varied.  Phenomenologically, $a$ and $b$ are not under
experimental control, so
$A$ must be varied in the appropriate range such that the data on $F_q$ are in
the range that exhibits $F$ scaling.  Now, on the lattice we must work with
specific bin sizes $\delta$.  That necessity destroys the advantage of dealing
with just one $x$ variable in the analytic method.  We make the identification 
\begin{eqnarray}
A/b = \delta^2
\label{18}
\end{eqnarray}
and choose a range of $\delta$ that can cover the scaling range in $x$ mentioned
above, i.e., $0.018< x < 7.5$.  The ratio of the extrema of that $x$ range
corresponds, for $a$ held  constant, to the ratio of the extrema of
$\delta$ being very nearly $20:1$.  Thus, if we let the smallest bin to consist
of just two sites in each direction, then the range of $\delta$ should be $2\leq
\delta \leq 40$.  Choosing the lattice size to be $L=120$, we have for the number of
bins, $M$, on the lattice to vary from $60^2$ down to $3^2$, a range that is wide
enough to test scaling.  The corresponding value of $a$ is roughly $-0.1$.

Applying the above parameters to (\ref{9}) and (\ref{10}), and using Monte Carlo
method to calculate $f_q$ and then $F_q$, we find the result in Fig.\ 1 shown in
log-log plot.  The linear dependences for $3\leq q \leq 6$ confirm the
$F$-scaling behavior.  Moreover, the scaling exponent $\nu$ defined in
(\ref{2}) is found to be $\nu =1.308$, as shown in Fig.\ 2, in essential agreement
with the result obtained in \cite{1}, as it should.

Proceeding now to the main task of including spatial fluctuations, we start by
setting $C=1$ in (\ref{12}) initially.  Other values of $C$ will be considered later.
We use the Metropolis algorithm to simulate the configurations of $z_i$ on each
site of the lattice, with the
Boltzmann factor $e^{-F_s}$ as the weight.  We discard the first $800$ sweeps for
initialization.  From
$10^3$ such configurations we calculate the distribution in $\zeta$.  However, owing to
(\ref{17}) it is more useful to plot the distribution as a function of
\ $\alpha \zeta$.  The arbitrarily chosen value of $\alpha$ affects only the scale of
the abscissa and the normalization of $S(\alpha \zeta)$.  Fig.\ 3 shows the results
for $\delta=2$ and $40$, and for $\alpha=1/4$, which is chosen so that for the
smallest bin the maximum multiplicity due to surface fluctuation is $\alpha \zeta=1$. 
As Fig.\ 3(a) shows, the distribution is very nearly Gaussian with half-width around
$0.3$, plus a sharp peak at $\alpha \zeta =0$, corresponding to the preponderant
probability that $z_i =0$ for all four sites in a bin.  On the other hand, for the
largest bin $\delta=40$, the distribution shown in Fig.\ 3(b) is negligible for
$\alpha \zeta >50$, although the maximum possible value, for all $z_i =+1$, is
$\alpha \zeta =40^2 /4 =400$. Apart from the peak at $\alpha \zeta =0$ the
half-width is around $20$.  The statistics is poorer in this case because the
lattice has far fewer bins per configuration when $\delta$ is increased twenty-fold.
As the area $\delta^2$ is increased by a factor of 400, the mean multiplicity
fluctuation is increased by a factor of only $20/0.3 \sim 60$.  Thus the deviation
from a flat surface does not have long wavelength modes, which would have resulted in
large $\zeta$ configurations in large bins.  On the other hand, the surface
fluctuations are not all local, lest Fig.\ 3(b) would not be as broad as it is.

Putting these $S(\alpha \zeta)$ distributions for various values of $\delta$
between $2$ and $40$ in (\ref{15}) and performing the convolution, we obtain the
dynamical multiplicity distribution $D(\mu)$.  Fig.\ 4 shows the results for a range
of values of $\delta$ with $C=1$ and $\alpha = \frac{1} {4}$.  From the $D(\mu)$
distribution we can use (\ref{6}) and (\ref{14}) to calculate $F_q$, which we present
in the log$F_q$ vs log$F_2$ form in Fig.\ 5.  We see that the straightline behavior
supports $F$ scaling.  As $\alpha$ is varied from $1/4$ to $2$, the ranges of $F_q$
values are reduced but the slopes of the straightlines are essentially unchanged.
Thus the first part of our objective has been accomplished; that is, $F$ scaling
persists even when surface fluctuations are considered.

Before we investigate the second part, {\it viz.}, the effect on scaling exponent, we
examine the dependence on $C$.  The foregoing analysis is done for $C=1$, a value
that has been arbitrarily chosen.  We now vary $C$ to see the effect on $S(\zeta)$. 
From the simulated result on $S(\zeta)$ for every set of values of $C$ and $\alpha$,
we calculate the width $\Gamma_\zeta = \left< \zeta^2 \right>^{1/2}$  and show in
Fig.\ 6 the dependence of $\Gamma_\zeta$ on $C$ for the two extreme bin-sizes,
$\delta=2$ and $40$.  We see that for $\delta =40$ there is a maximum at $C=2$, so
the previous result on $C=1$ is not far from maximum surface fluctuation. 
$\Gamma_\zeta$ is lower at both higher and lower values of $C$ because high $C$
stiffens the surface and suppresses $\zeta$, while low $C$ leads to random $z_i$ and
therefore also reduces $\zeta$.  The same is not true for $\delta =2$, since even in
the limit of $C\rightarrow 0$ the smallness of the bin size makes the cancellation of
random $z_i$ ineffective, resulting in a Possionian $S(\zeta)$ that has nonvanishing
width.  As $C$ increases, the surface becomes smoother, and $\Gamma_\zeta$ continues
to decrease monotonically.  To summarize, we need only consider the range $0<C<2$ for
maximum effect due to surface fluctuation.  

We now can repeat the previous calculations for other values of $C$ and produce
figures similar to those shown in Figs.\ 4 and 5.  However, since the results are
very similar, we omit those figures.

To determine the slopes $\beta_q$ of the $F$-scaling behavior, we fit all the points
in Fig.\ 5 for a given $q$ value, including all values of $\alpha$, by a straight
line.  The resultant $\beta_q$ is then shown in Fig.\ 7 for a particular value of \ 
$C$ ($C=1$ for Fig.\ 5).  The error bars in Fig.\ 7 correspond to the deviations from 
the straightline fits of the points in Fig.\ 5  for $C=1$ only.  For other values of
$C$ the results from similar calculations are also shown in Fig.\ 6.  No corresponding
error bars are shown, since they are similar.  The $\beta_q$ values are determined
only for integer values of $q$, but since the dependence on $C$ is so insignificant,
we have for clarity exhibited the $\beta_q$ values by straightline fits according to
(\ref{2}).  Even so, the overlap obsures the individual lines for different values of
$C$, but successfully demonstrates that the scaling exponent $\nu$ is insensitive to
$C$.  The net value of $\nu$ that can be concluded from Fig.\ 7 is 
\begin{eqnarray}
\nu = 1.306 \pm 0.035  ,
\label{19}
\end{eqnarray}  
when surface fluctuations are taken into account.  Compared to the
result of $\nu= 1.308$ obtained from Fig.\ 2 for pure GL, we see
that the change in $\nu$ is well within the errors.  

\section {Conclusion}

To study the effect of surface fluctuation on the observables of
quark-hadron PT, we have incorporated into the GL description an additional
term in the free energy that depends on the vertical displacements from a
flat surface.  We have used lattice simulation to determine the probability
that a bin has a net displacement
$\zeta$.  The physics that controls such displacements is color confinement.  We
connect the surface fluctuations to multiplicity fluctuations, which in turn are
quantified by normalized factorial moments $F_q$.  The $F$ scaling of $F_q$, when
the bin size is varied, results in the determination of the scaling exponent
$\nu$.  The aim of this work has been to study the dependence of $\nu$ on surface
fluctuations, after verifying that the scaling behavior is not destroyed.    

There are two parameters in the calculation, on which we have no {\it a priori}
information.  One is $C$; the other is $\alpha$.  $C$ depends on the strength of
confinement, and relates the vertical displacement $z_i$ (arbitrarily normalized)
to the free energy (normalized by the transition temperature).  The parameter
$\alpha$ relates the net displacement $\zeta$ of a cell to the deviation of the
cell multiplicity from the mean.  We have studied the variation of $\nu$, when $C$
and $\alpha$ are varied over the important ranges $0< C <2$ and $0< \alpha <2$.  When
$C \simeq 2$ or $\alpha \simeq 2$, we do see significant fluctuations.  Our result,
however, shows that $\nu$ is essentially unchanged when $C$ and $\alpha$ are
varied in the ranges considered.  There is no doubt that when $\alpha$ is large
enough, the scaling behavior will break down.  But then our method of calculation
should become invalid in that case, since large multiplicity fluctuations cannot
be treated in the framework of the mean-field theory.  
Besides, there is no physical reason to expect that small surface fluctuation
can lead to large multiplicity fluctuations.  

The significance of this
work is therefore in finding that the scaling behavior of pure GL is stable
against the perturbation introduced by surface fluctuations.  Thus this result
improves the Ginzburg-Landau description of quark-hadron phase transition and
provides a more physical connection to the realistic situation in heavy-ion
collisions.  The end result is that $F$ scaling persists even when surface
fluctuations are taken into account, and that the scaling exponent, which is
experimentally measurable, remains at $\nu \simeq 1.3$.

\begin{center}

{\large {\bf Acknowledgment}}
\end{center}

RCH would like to thank Murat Nazirov for discussions during the initial
stage of this investigation.  This work was supported, in part, by the U.S.
Department of Energy under grant No. DE-FG06-91ER40637.

\vfill
\newpage
\centerline{\large \bf Figure Captions}
\vskip.5cm
\begin{description}
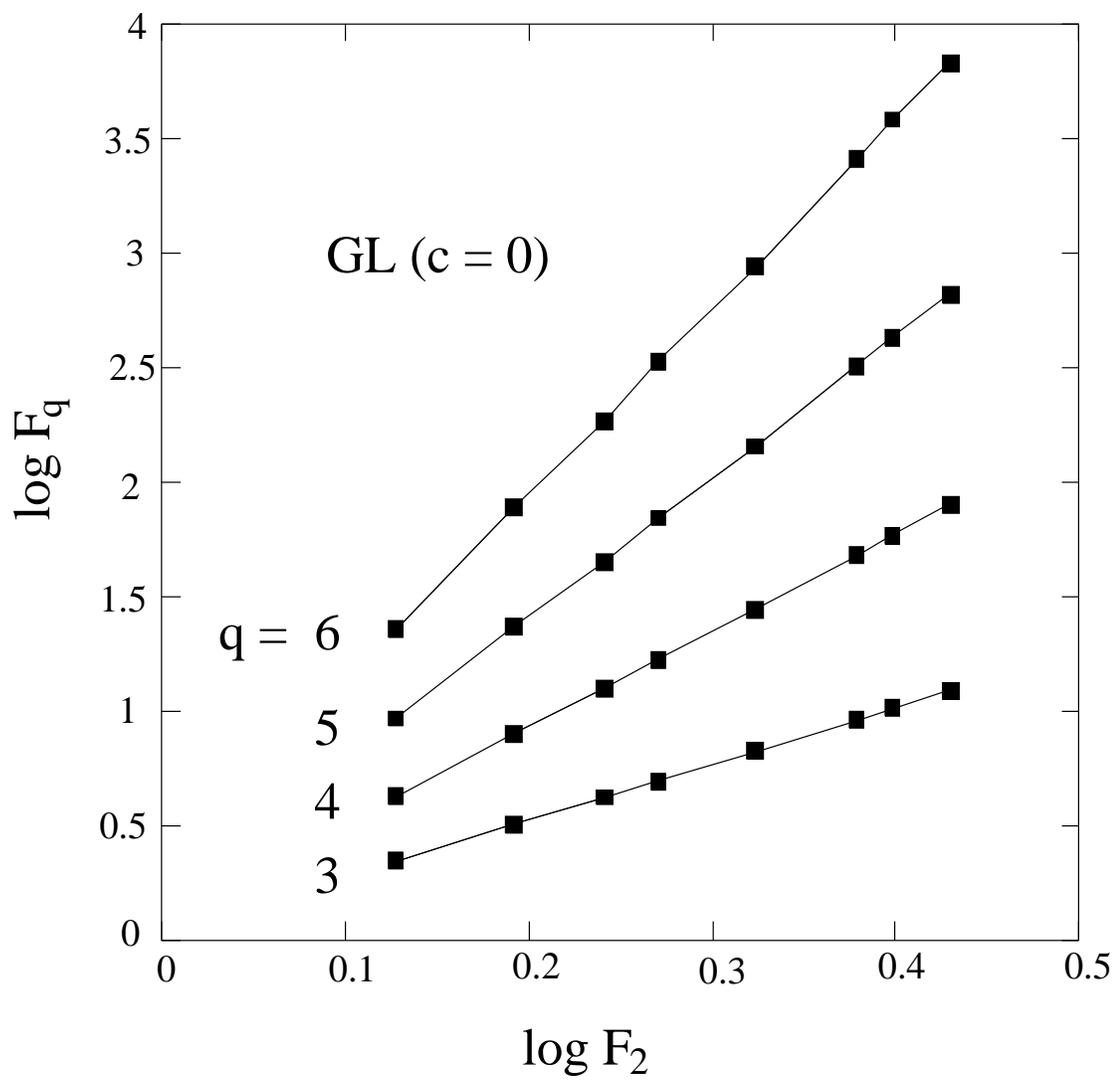
 \item[Fig.\ 1 \quad] $F$ scaling in Ginzberg-Landau theory without spatial
inhomogeneity, i.e., $c=0$ in Eq.\ (4).

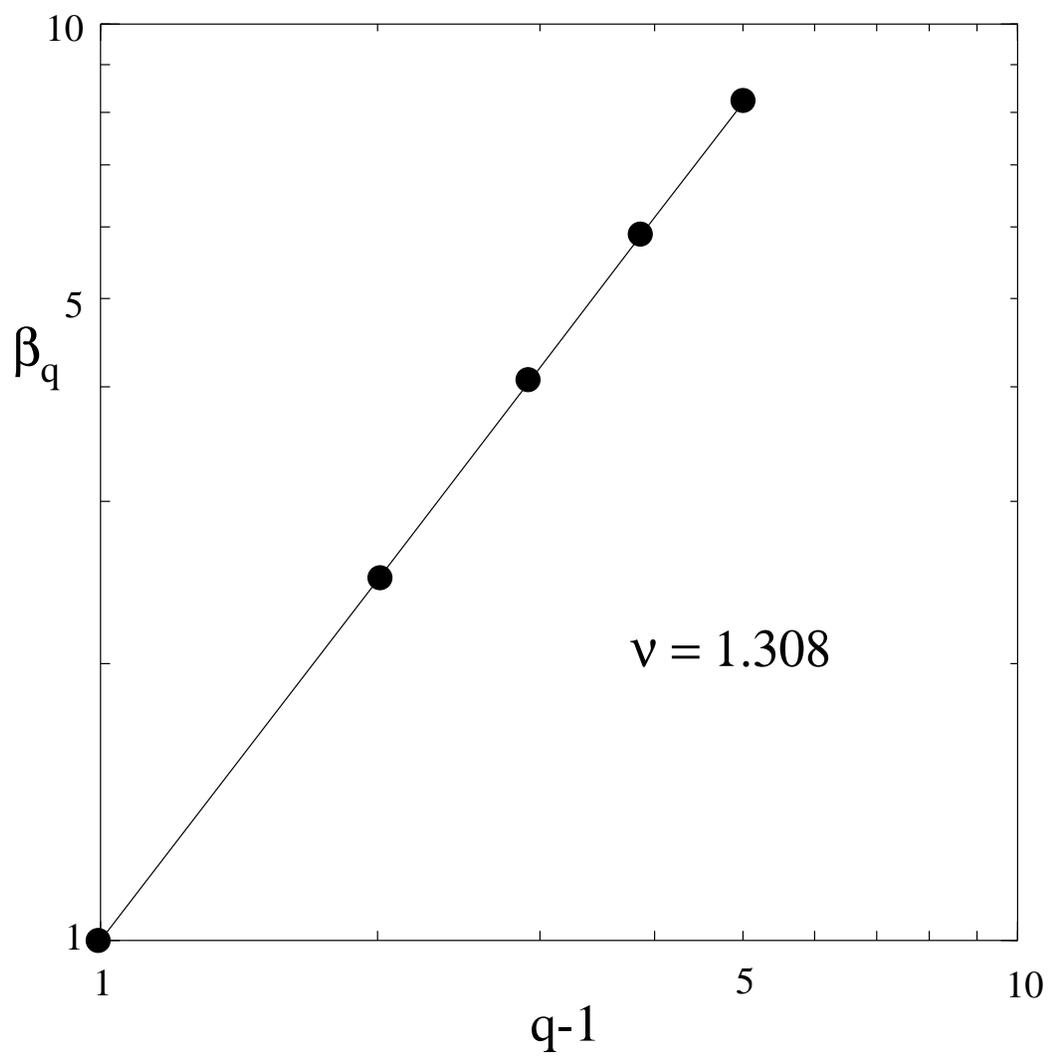
\item[Fig.\ 2  \quad] Log-log plot verifying Eq.\ (2), using the data of Fig.\ 1.

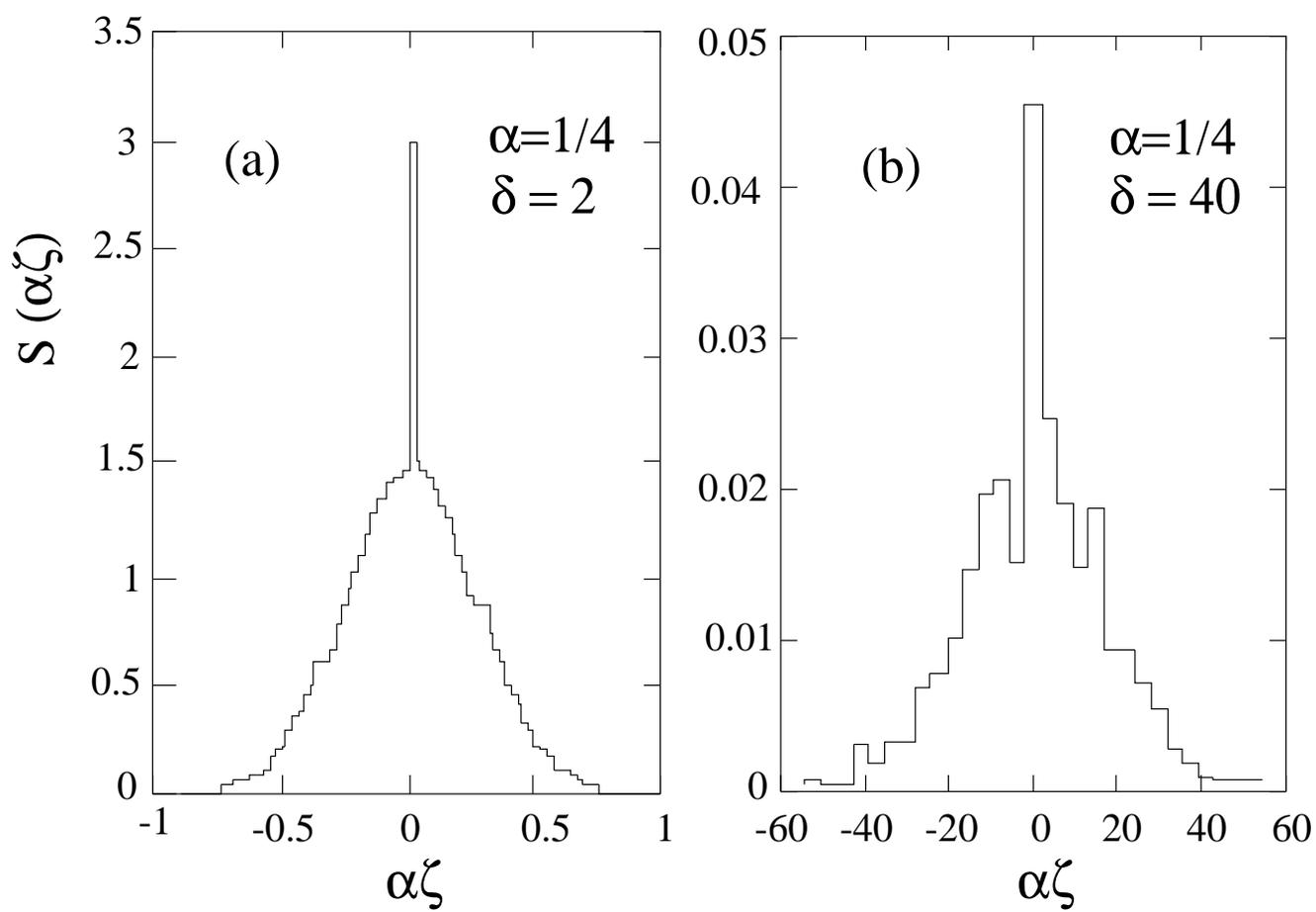
\item[Fig.\ 3  \quad] Distribution of surface fluctuation for $C=1$ and $\alpha =1/4$. 
The bin sizes are different in (a) and (b) as indicated.\

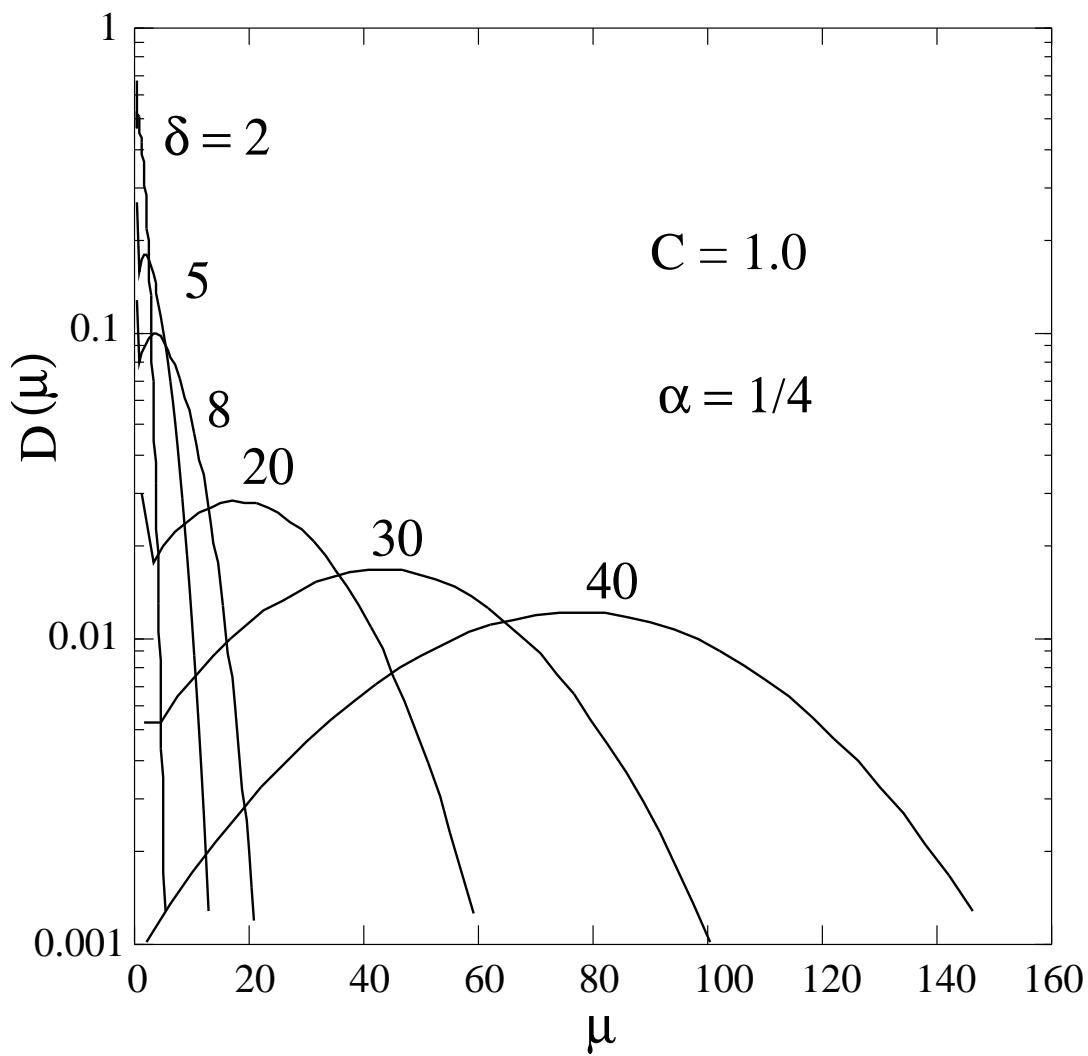
\item[Fig.\ 4  \quad] Distribution of dynamical multiplicity when the GL description of
PT is supplemented by surface fluctuations.

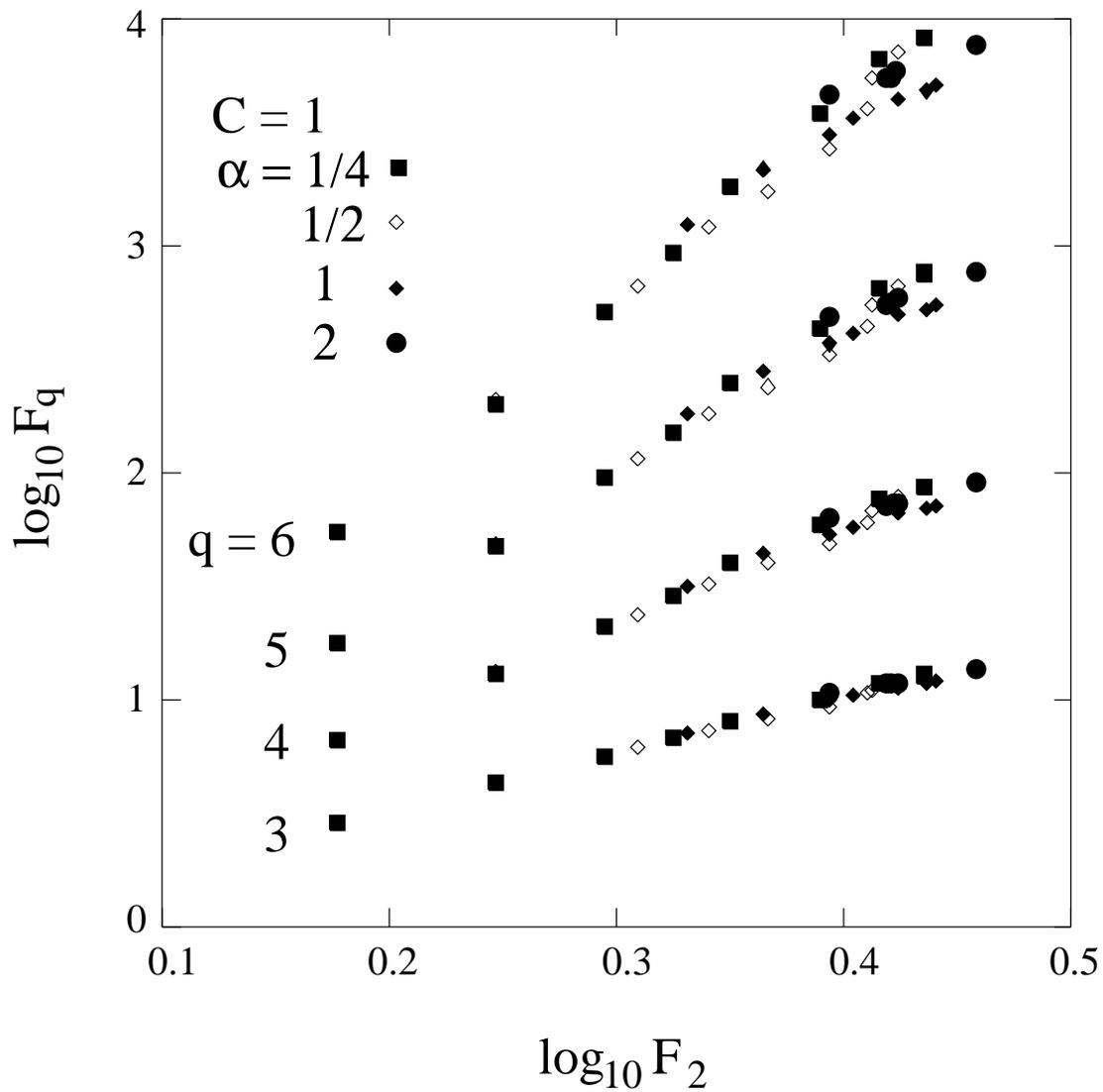
\item[Fig.\ 5  \quad] $F$-scaling plot of $F_q$ for $C=1$ and a range of $\alpha$.

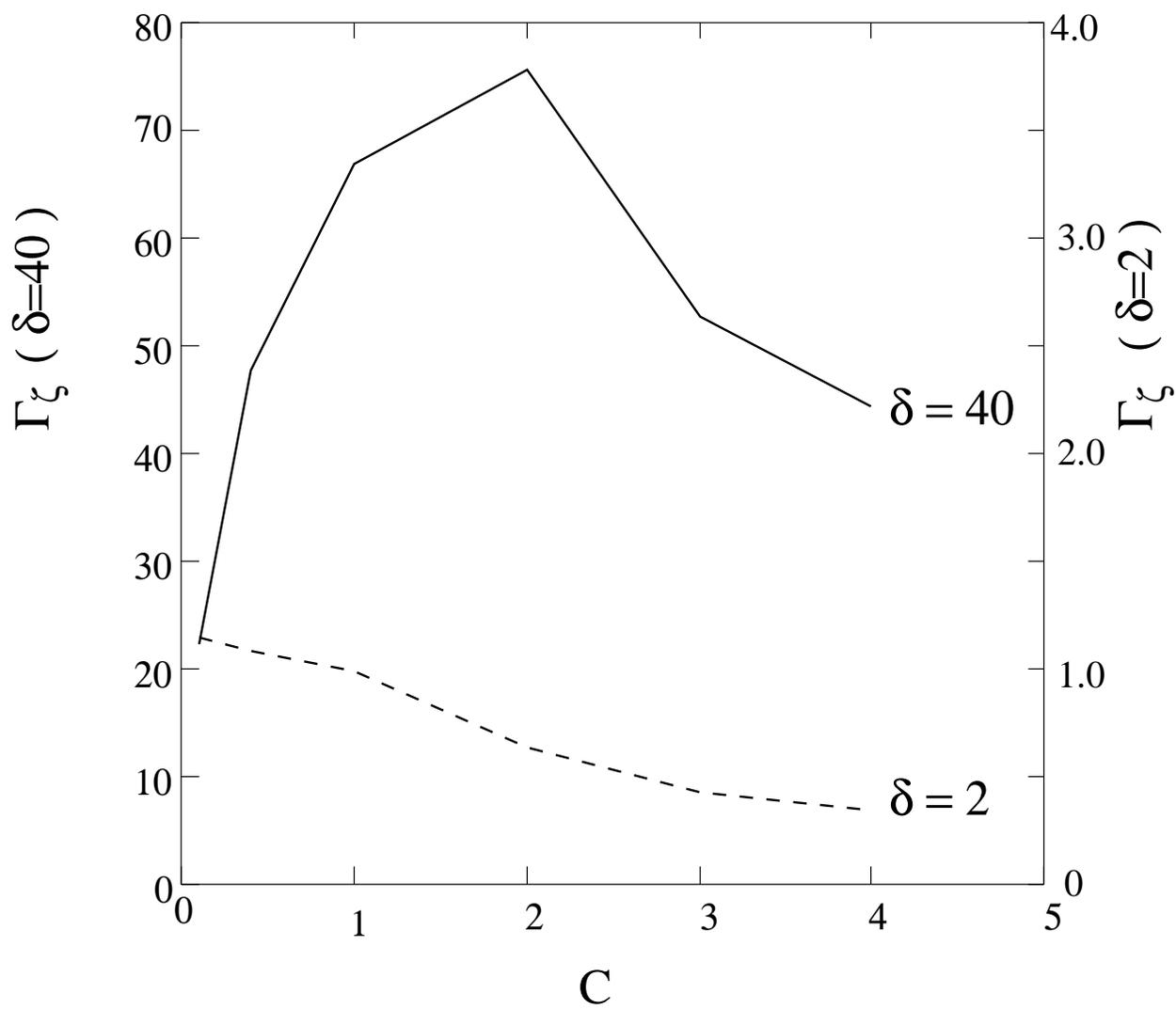
\item[Fig.\ 6  \quad] Width of $S(\zeta)$ vs $C$ for $\delta =2$ (right scale) and
$40$ (left scale).

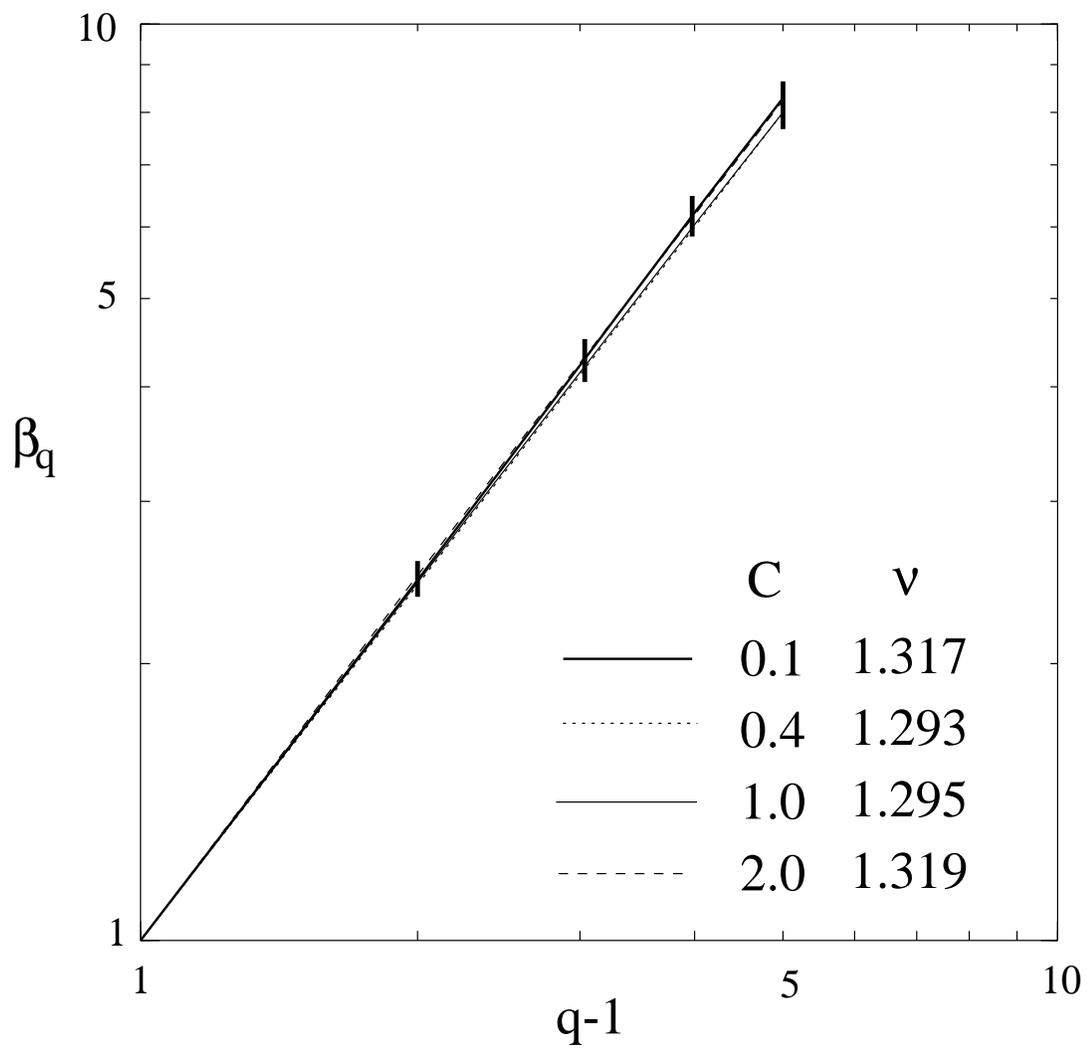
\item[Fig.\ 7  \quad] The slopes $\beta_q$ for a range of values of $C$.  Although
$\beta_q$ are determined for integer values of $q$, they are plotted as straight lines
to avoid overlap of points.  The error bars are for $C=1$ only, the ones for other
values of
$C$ being similar.

\end{description}
\end{document}